\documentclass[a4paper]{jpconf}
\usepackage{graphicx}
\usepackage{cite}

\usepackage{amsmath,amsfonts,amssymb}

\newcommand{\bo}{\raise-1mm\hbox{\Large$\Box$}} 

\newcommand{\bra}[1]{\langle #1|} 
\newcommand{\ket}[1]{|#1\rangle}

\newcommand{\eb}{\begin{equation}} 
\newcommand{\ee}{\end{equation}}

\newcommand{\lt}{\left}
\newcommand{\rt}{\right}

\newcommand{\tev}{\,\mbox{TeV}}
\newcommand{\gev}{\,\mbox{GeV}}
\newcommand{\mev}{\,\mbox{MeV}}
\newcommand{\Kbar}{\,\overline{\!K}}
\newcommand{\Dbar}{\,\overline{\!D}}
\newcommand{\Bbar}{\,\overline{\!B}}

\newcommand{\bbs}{$\mathrm{B_s}\!-\!\ov{\mathrm{B}}{}_\mathrm{s}\,$}
\newcommand{\bbms}{$\mathrm{B_s}\!-\!\ov{\mathrm{B}}{}_\mathrm{s}\,$\ mixing}
\newcommand{\bbmd}{$\mathrm{B_d}\!-\!\ov{\mathrm{B}}{}_\mathrm{d}\,$\ mixing}
\newcommand{\bbmq}{$\mathrm{B_q}\!-\!\ov{\mathrm{B}}{}_\mathrm{q}\,$\ mixing}

\newcommand{\bbm}{$\mathrm{B}\!-\!\ov{\mathrm{B}}{}\,$\ mixing}

\newcommand{\dm}{\ensuremath{\Delta m}}
\newcommand{\dg}{\ensuremath{\Delta \Gamma}}

\newcommand{\BdorBdbar}{\raisebox{7.7pt}{$\scriptscriptstyle(\hspace*{8.5pt})$}
  \hspace*{-10.7pt}\!\Bbar_{d}}

\newcommand{\bea}{\begin{eqnarray}}
\newcommand{\eea}{\end{eqnarray}}

\newcommand{\nn}{\nonumber \\}
\newcommand{\no}{\nonumber}
\newcommand{\ov}{\overline}
\newcommand{\epm}[2]{
 \raisebox{-0.5ex}{\shortstack[l]{$\scriptstyle+#1$\\$\scriptstyle-#2$}}}

\newcommand{\eq}[1]{Eq.~(\ref{#1})}
\newcommand{\eqsand}[2]{Eqs.~(\ref{#1}) and (\ref{#2})}
\newcommand{\fig}[1]{Fig.~\ref{#1}}

\begin{document}
\title{Flavour and CP Violation}

\author{Ulrich Nierste}

\address{Institut f\"ur Theoretische Teilchenophysik\\ 
Karlsruhe Institute of Technology\\ Engesserstraße 7\\ 
76131 Karlsruhe, Germany}

\ead{ulrich.nierste@kit.edu}

\begin{abstract}
In this talk I review the status of \bbm\ in the Standard Model and the
room for new physics in \bbs\ and \bbmd\ in the light of recent LHCb
data. 
\end{abstract}

\section{Introduction}
The only source of flavour-changing transition in the Standard Model
(SM) is the Yukawa interaction of the Higgs doublet $H$ with the fermion
fields.  The Yukawa lagrangian of the quark fields reads
\begin{eqnarray}
-L_Y = Y_{jk}^d\, \ov Q{}^{\,j}_L \,H\, d_R^{\,k} \;+\; 
            Y_{jk}^u\,\ov Q{}^{\,j}_L \,\widetilde H \,u_R^{\,k} 
     \;+\; \mbox{h.c.}
\end{eqnarray}
Here $Q{}^{\,j}_L$ denotes the SU(2) doublet of the left-handed quark
fields of the $j$-th generation and $d_R^{\,j}$, $u_R^{\,j}$ are the
corresponding right-handed singlet fields. The $3\times 3$ Yukawa
matrices $Y^{d,u}$ and the Higgs field vacuum expectation value
$v=174\,\gev$ combine to the quark mass matrices $M^{d,u}=Y^{d,u} v$. 
The gauge interactions of the quark fields do not change under 
unitary rotations of any of $Q{}^{\,j}_L$,  $d_R^{\,j}$, and $u_R^{\,j}$ 
in flavour space. Using these unphysical transformations one can 
bring $Y^u$ and $Y^d$ to the form 
\begin{eqnarray}
  Y^u= \widehat Y^u = \begin{pmatrix} y_u&0&0\\ 0&y_c&0 \\ 0&0&y_t 
       \end{pmatrix} \qquad \mbox{and}\qquad 
     Y^d= V^\dagger \widehat Y^d && \quad \mbox{with}
     \quad \widehat Y^d = \begin{pmatrix} y_d&0&0\\ 0&y_s&0 \\ 0&0&y_b 
       \end{pmatrix}
  \label{bas}
\end{eqnarray}
and $y_i\geq 0$. With the choice adopted in \eq{bas} the mass matrix
$M^u$ of up, charm, and top quark is diagonal. The diagonalisation of
$M^d$ requires the additional rotation $d_L^{\,j} = V_{jk}
d_L^{\,k\prime}$, which puts the unitary Cabbibbo-Kobayashi-Maskawa
(CKM) matrix $V$ into the $W$ boson vertices: $W_\mu \ov u_L^{\,j}
\gamma^\mu d_L^{\,j} = W_\mu V_{jk} \ov u_L^{\,j} \gamma^\mu
d_L^{\,k\prime}$.  \eq{bas} defines a basis of weak eigenstates which is
particularly suited to display the smallness of flavour violation in the
SM. All flavour-changing transitions originate from $Y^d$ which reads 
\begin{eqnarray}
Y^d= V^\dagger \widehat Y^d &=& 
    \begin{pmatrix} 
    10^{-5}& -7\cdot 10^{-5} & (12 + 6 i)\cdot 10^{-5} \\
    4\cdot 10^{-6} & 3 \cdot 10^{-4} & -6\cdot 10^{-4} \\  
    (2+6 i) \cdot 10^{-8} & 10^{-5} &  2 \cdot 10^{-2} 
    \end{pmatrix},
\end{eqnarray}
if all Yukawa couplings are evaluated at the scale $\mu=m_t$.  The
off-diagonal element largest in magnitude is $V_{ts}^* y_b\equiv
  V_{32}^* y_b=-6\cdot 10^{-4}$. Moreover, in the SM flavour-changing
neutral current (FCNC) transitions involve an additional loop
suppression, making FCNC processes an extremely sensitive probe of
physics beyond the SM. This feature puts flavour physics into a
win-win position: If CMS and ATLAS find new particles, FCNC observables
will be used to probe the flavour pattern of the BSM theory to which
these particles belong.  If instead CMS and ATLAS do not find any BSM
particles, flavour physics will indirectly probe new physics to scales
exceeding 100$\,\tev$, well beyond the center-of-mass energy of the LHC.

\begin{figure} 
\centerline{
\parbox{0.4\textwidth}
{\includegraphics[height=3cm]{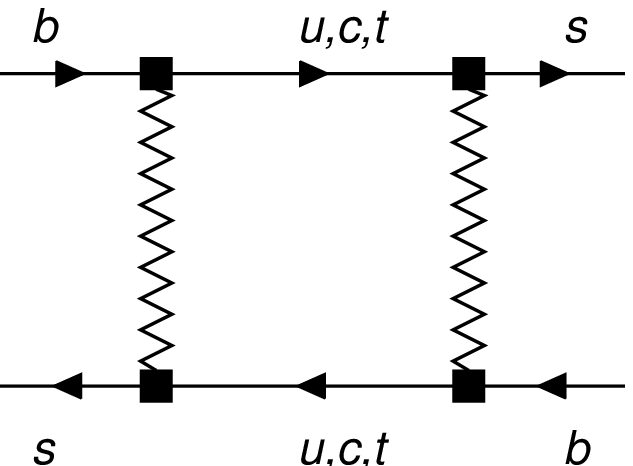}}~
\parbox{0.4\textwidth}{\flushright
\includegraphics[height=2.5cm,angle=-90]{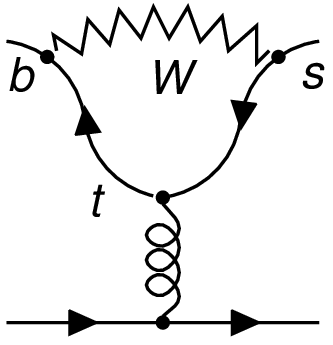}}}
\caption{Left: SM box diagram describing \bbms, a $|\Delta B|=2$
  process. Right: $b\to s$ penguin amplitude, a  $|\Delta B|=1$ 
transition.\label{fig:box} 
}
\end{figure} 
The weak interaction makes the neutral $K \sim \ov s d$, $D\sim c \ov
u$, $B_d\sim \ov b d$ and $B_s\sim \ov b s$ mesons mix with their
antiparticles, $\Kbar$, $\Dbar$, $\Bbar_d$, and $\Bbar_s$,
respectively. This means that, say, a $B_d$ meson evolves in time into a
quantum-mechanical superposition of a $B_d$ and a $ \Bbar_d$. This
feature is a gold-mine for new-physics searches, as it permits to probe
CP phases in any decay to a final state which is accessible from both
the $B_d$ and the $ \Bbar_d$ component of the decaying state. In a
tree-level decay like $\BdorBdbar \to J/\psi K_S$ the time-dependent CP
asymmetry probes new physics in the mixing amplitude itself. If instead
a rare decay (with FCNC decay amplitude) is studied, the time-dependent
CP asymmetry may reveal a new physics contribution to the decay
amplitude. Meson-antimeson mixing is a $|\Delta F|=2$ process, meaning
that the relevant flavour quantum numbers change by two units. In the
case of \bbms\ depicted in \fig{fig:box} these flavour quantum numbers
are beauty $B$ and strangeness $S$. Weak decays are $|\Delta F|=1$
processes, \fig{fig:box} shows a $|\Delta B|=|\Delta S|=1$ FCNC decay
amplitude. It is instructive to compare the reach of $|\Delta F|=2$ and
$|\Delta F|=1$ amplitudes to new physics: The flavour violation in the
SM the amplitudes $A^{|\Delta B|=2}_{\rm SM}$ and $A^{|\Delta B|=1}_{\rm
  SM}$ in \fig{fig:box} is governed by the small CKM element $V_{ts}$
and both diagrams scale as $1/M_W^2$. A new-physics (NP) contribution
may instead involve some new flavour-violating parameter $\delta_{\rm
  FCNC}$. If the scale of NP is $\Lambda$ and the NP contribution enters
at the one-loop level, one finds 
\begin{eqnarray}
\frac{|A^{|\Delta B|=2}_{\rm NP} |}{|A^{|\Delta B|=2}_{\rm SM}|} = 
     \frac{|\delta_{\rm FCNC}|^2}{|V_{ts}|^2} \frac{M_W^2}{\Lambda^2} 
\qquad&\mbox{and}& \qquad 
\frac{|A^{|\Delta B|=1}_{\rm NP}|}{|A^{|\Delta B|=1}_{\rm SM}|}=
     \frac{|\delta_{\rm FCNC}|}{|V_{ts}|} \frac{M_W^2}{\Lambda^2}.
\label{smnp}
\end{eqnarray}
One sees that \bbms\ is more sensitive to generic NP than an FCNC $b\to
s$ decay amplitude, if $|\delta_{\rm FCNC}|>|V_{ts}|\approx 0.04$. The
estimate in \eq{smnp} applies, for example, to the (N)MSSM with generic
flavour structure in the bilinear SUSY-breaking terms, which induce new
$b_R\to s_R$ or $b_L\to s_L$ transitions through squark-gluino loops
\cite{Gabbiani:1996hi,Virto:2009wm}.  But $|\Delta B|=1$ transitions can
be more sensitive probes of NP in models in which $b_R\to s_L$ or
$b_L\to s_R$ transitions are parametrically enhanced over their SM
counterparts. This situation occurs in the (N)MSSM with a large value of
the parameter $\tan\beta$ or large trilinear SUSY-breaking
terms. Theories with such chirally enhanced FCNC transitions are
efficiently probed with the radiative decays $b\to s \gamma, d\gamma$
\cite{Bertolini:1990if,Oshimo:1992zd,Carena:2000uj,Degrassi:2000qf,
  Buras:2002vd,Baer:1997jq,Crivellin:2009ar,Crivellin:2011ba}, leptonic
decays like the recently observed \cite{Aaij:2012nna} decay $B_s\to
\mu^+ \mu^-$ \cite{Babu:1999hn,Huang:2000sm,Dedes:2001fv,Isidori:2001fv,
  Buras:2002wq,Buras:2002vd,Gorbahn:2009pp,Hofer:2009xb,Altmannshofer:2009ne,
  Crivellin:2010er,Altmannshofer:2012ks} and the chromomagnetic
contribution to non-leptonic decays like $B_d\to \phi K_S$
\cite{Hofer:2009xb}.

\section{\boldmath \bbm: formalism and Standard-Model prediction}
\bbmq\ with $q=d$ or $q=s$ is governed by the $2\times 2$ matrix $M^q-i
\Gamma^q/2$, with the hermitian mass and decay matrices $M^q$ and
$\Gamma^q$. Due to non-vanishing off-diagonal elements $M^q_{12}$ and
$\Gamma^q_{12}$ a meson tagged as a $B_q$ at time $t=0$ will for $t>0$
evolve into a superposition of $B_q$ and $\Bbar_q$.  In the SM $
M_{12}^s$ is calculated from the dispersive part of the box diagram in
\fig{fig:box}, which amounts to discarding the imaginary part of the
loop integral. $\Gamma^s_{12}$ is instead obtained from the absorptive
part, meaning that the real part of the loop integral is dropped. 
$M_{12}^d$ and $\Gamma^d_{12}$ are calculated in the same way from
the box diagram with the external $s$ quarks replaced by $d$ quarks. 
 $M^q_{12}$ is dominated by the top contribution, while $\Gamma_{12}^q$ 
stems from the diagrams with internal up and charm.  $\Gamma_{12}^q$ is made up
of all final states into which both $B_q$ and $\Bbar_q$ can decay.   
There are three physical quantities in {\bbmq}:
\begin{displaymath}
\lt| M_{12}^q \rt|,\quad  \lt| \Gamma_{12}^q \rt|,\quad
 \phi_q\equiv \arg \lt( - \frac{M_{12}^q}{\Gamma_{12}^q} \rt) .
\end{displaymath}
By diagonalising $M-i\,\Gamma/2$ one finds the two 
mass eigenstates $B_q^H$ and $B_q^L$, where the superscripts stand for
``heavy'' and ``light''. These eigenstates differ in their mass and
width with 
\begin{eqnarray}
\dm_q & = & M_H^q-M_L^q \; \simeq \; 2 |M_{12}^q| \, , \qquad\qquad
\dg_q \;=\; \Gamma_L^q-\Gamma_H^q \;\simeq \; 2 |\Gamma_{12}^q| \cos
\phi_q  \, .\no
\end{eqnarray}
The mass difference $\dm_q$ simply equals the frequency at which $B_q$
and $\Bbar_q$ oscillate into each other. The width difference $\dg_s$ is
sizable, so that in general untagged $B_s$ decays are governed by the
sum of two exponentials. There is no useful data on the tiny width
difference $\dg_d$ in the $B_d$ system yet, which is predicted to be
around $0.5\%$ of $\Gamma_{L,H}^d$
\cite{Beneke:2003az,Ciuchini:2003ww,Lenz:2006hd,ckm10}. The CP-violating
phase $\phi_q$ can be determined by measuring the CP
asymmetry in flavour-specific decays: 
\begin{eqnarray}
a_{\rm fs}^{q} &=& 
\frac{|\Gamma_{12}^q|}{|M_{12}^{q}|} \sin \phi_q \, .\no
\end{eqnarray}
A decay $B_q \to f$ is called flavour-specific, if the decay $\Bbar_q \to f$ is
forbidden. The standard method to determine $a_{\rm fs}^{q}$ uses
semileptonic decays, so that $a_{\rm fs}^{q}$ is often called
semileptonic CP asymmetry. While this measurement simply requires to
count the numbers of positive and negative leptons in $B_q$ decays, it
is still very difficult, because $|\Gamma_{12}^q| \ll |M_{12}^{q}|$ renders
$|a_{\rm fs}^{q}|$ small, even if NP contributions to $M_{12}^q$ enhance
$\phi_q$ over its small SM value.

\subsection{Mass difference $\dm_s$}
\begin{figure} 
\centerline{
\includegraphics[height=2.5cm]{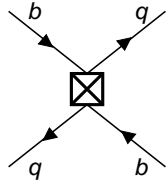}}
\caption{The $|\Delta B|=2$ operator $Q$ of \eq{q}. It is pictorially
  obtained by shrinking the \bbmq\ box diagram to a point.\label{fig:op}
}
\end{figure}

The theoretical prediction of $\dm_q$ requires the separation of 
short-distance and long-distance QCD effects. To this end one employs an
operator product expansion, which factorises $M_{12}^q$ as 
\begin{eqnarray}
M_{12}^q = \lt( V_{tq}^* V_{tb} \rt)^2 \, C \bra{B_q} Q \ket{\Bbar_q}. 
   \label{m12}
\end{eqnarray}
The Wilson coefficient $C$ comprises the short-distance physics, 
with all dependence on the heavy particle masses. QCD 
corrections to $C$ have been calculated reliably in perturbation 
theory \cite{bjlw}. 
Since the CKM elements are factored out in \eq{m12},
$C=C(m_t,M_W,\alpha_s)$ is real in the SM. 
\begin{eqnarray}
 Q & =& 
    \ov{q}_L \gamma_{\nu} b_L \; \ov{q}_L \gamma^{\nu} b_L  \label{q}
\end{eqnarray}
is a local four-quark operator describing a point-like $|\Delta B|=2$ 
transition, see \fig{fig:op}. 
The formalism is the same for the $B_d$ and $B_s$ complex; for
definiteness I only discuss the case $q=s$ in the following. The
average  of the CDF and LHCb measurements of $\dm_s\simeq 2|M_{12}^s|$ is 
\cite{hfag}
\begin{eqnarray}
  \dm_s^{\rm exp}= (17.719 \pm 0.043 )\, \mbox{ps}^{-1},  \label{dmsexp}
\end{eqnarray}
meaning that $|M_{12}^s|$ is known very precisely. However, we need the
hadronic matrix element $ \bra{B_s} Q \ket{\Bbar_s}$ to confront
\eq{dmsexp} with the SM. The latter is usually parametrised as
\begin{eqnarray}
\bra{B_s} Q \ket{\Bbar_s} &=& 
  \frac{2}{3} M_{B_s}^2 \,  f_{B_s}^2 \, B_{B_s} . \no
\end{eqnarray}
Here $M_{B_s}$ and $ f_{B_s}$ are mass and decay constant of the $B_s$
meson, respectively. The hadronic parameter $ B_{B_s}$ depends on the
renormalisation scheme and scale used to define $Q$, in this talk $
B_{B_s}\approx 0.85$ is understood to be evaluated at the scale
$\mu=m_b$ in the $\ov{\rm MS}$ scheme. Noting that $|V_{ts}|$ is fixed
by CKM unitarity from the well-measured element $|V_{cb}|$, we can write
\begin{eqnarray}
\dm_s = \lt( \lt.\lt.\lt. 
           18.8 \pm 0.6\rt._{V_{cb}} \pm 0.3\rt._{m_t} 
                \pm 0.1\rt._{\alpha_s} \rt) \, \mbox{ps}^{-1}\, 
     \frac{f_{B_s}^2 \, B_{B_s} }{(220\mev)^2}, \label{dms}
\end{eqnarray}
where the  uncertainties from the experimental errors of the input
parameters indicated. Averaging various calculations from lattice gauge
theory one finds \cite{ckmf} 
\begin{eqnarray}
f_{B_s}^2 B_{B_s} = \lt[ (211 \pm 9) \mev \rt]^2 . \label{fbnum}
\end{eqnarray} 
Inserting this result into \eq{dms} implies 
\begin{eqnarray}
 \dm_s= (17.3 \pm 1.5 ) \, \mbox{ps}^{-1} \label{dmth} 
\end{eqnarray}
complying excellently with the experimental result in \eq{dmsexp}.
However, the average in \eq{fbnum} mainly involves different 
calculations of $f_{B_s}$, which are combined with two fairly old  
results for $B_{B_s}$. With the recent  preliminary Fermilab/MILC result 
\cite{Bouchard:2011xj}
\begin{eqnarray}
f_{B_s}^2 \, B_{B_s}=0.0559(68)\gev^2\simeq \lt[(237\pm 14) \mev\rt]^2
\no
\end{eqnarray}
one finds instead 
\begin{eqnarray}
\dm_s= (21.7 \pm 2.6 ) \, \mbox{ps}^{-1} .
\end{eqnarray}
Therefore more effort on lattice-QCD calculations of $f_{B_s}^2 \,
B_{B_s}$ is highly desirable. As long as modern calculations of this
quantity are unavailable, I recommend to inflate the error in
\eq{dmth} to $2.5\,\mbox{ps}^{-1}$. One concludes that the precise
measurement in \eq{dmsexp} still permits a NP contribution of $15\% $ in
$\dm_s$.

\subsection{CP phase from $B_s \to J/\psi \phi$}
\begin{figure} 
\centerline{
\includegraphics[clip,scale=0.7,viewport=100 540 530 700]{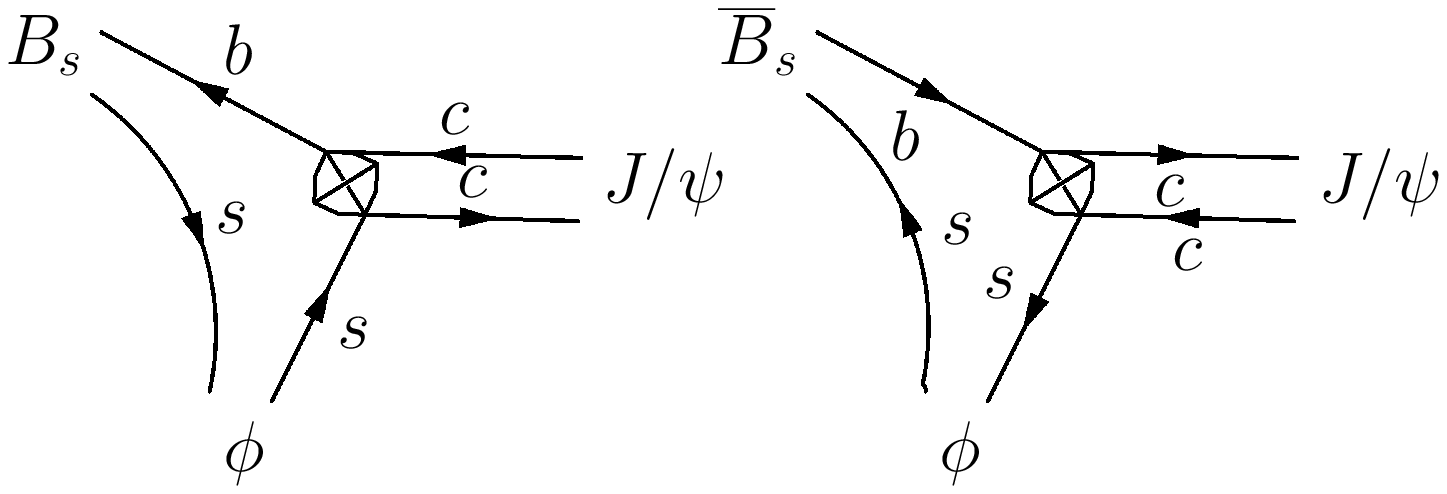}}
\caption{Amplitudes of $B_s \rightarrow J/\psi \,\phi$ and
  $\Bbar{}_s \rightarrow J/\psi \,\phi$. The cross denotes the 
$W$-mediated $b\to c \ov c s$ decay.\label{fig:psi}}
\end{figure}

While $\dm_s$ fixes the magnitude of $M_{12}^s$, the phase of $M_{12}^s$
can be probed through the mixing-induced CP asymmetry in $B_s \to J/\psi
\phi$. The final state contains two vector mesons, so that the orbital
angular momentum $L$ can assume the values $0$,$1$, and $2$. The final
states $(J/\psi \,\phi)_{L = 0,2}$ have CP quantum number $\eta_{CP}=1$,
while $(J/\psi \,\phi)_{L = 1}$ is CP-odd. Denoting a meson which was a
$B_s$ at time $t=0$ by $B_s(t)$ (with an analogous definition of
$\Bbar_s(t)$), one can define the time-dependent CP asymmetry
\begin{eqnarray}
a_{CP}(t) = 
         \frac{  \Gamma (B_s(t) \rightarrow (J/\psi \,\phi)_L  ) -
                 \Gamma (\Bbar{}_s(t) \rightarrow (J/\psi \,\phi)_L )}
              {  \Gamma (B_s(t) \rightarrow (J/\psi \,\phi)_L  ) +
                 \Gamma (\Bbar{}_s(t) \rightarrow (J/\psi \,\phi)_L )}.
\label{acpt}
\end{eqnarray} 
The two interfering amplitudes giving rise to $a_{CP}(t)$ are shown in
\fig{fig:psi}. 
The analytical result reads
\begin{eqnarray}
a_{CP}(t) = \eta_{CP} \frac{\sin(2\beta_s) \sin ({\dm_s} t)}{ \cosh
      ({\dg_s}t/2) - \cos (2\beta_s) \sinh ({\dg_s} t/2)}.
\label{acpt2}
\end{eqnarray}
The CP phase entering $a_{CP}(t)$ is $ \beta_s=\arg(-V_{tb}^{*}V_{ts}/
(V_{cb}^{*} V_{cs} ))$, in the standard phase convention of the CKM
matrix $ \beta_s$ is essentially just the phase of $-V_{ts}$. 
It should be mentioned that \eq{acpt} is a theoretical definition, in 
practice the experimental separation of the $L=0,1,2$ amplitudes 
involves a complicated angular analysis of the $J/\psi$ and $\phi$ 
decays. 

The CKM matrix can be parametrised in terms of the four Wolfenstein
parameters $\lambda$,$A$,$\ov{\rho}$, and $\ov{\eta}$
\cite{Wolfenstein:1983yz,Buras:1994ec}. Expanding to leading order in
$\lambda=0.225$ one finds $\sin(2\beta_s)\simeq 2 \lambda^2
\ov{\eta}$. The parameter $\ov{\eta}$ is the height of the CKM unitarity
triangle defined by $\ov{\rho} + i \ov{\eta} = - V_{ub}^*
V_{ud}/(V_{cb}^* V_{cd})$. The suppression by $\lambda^2$ renders
$\sin(2\beta_s)$ small; determining the CKM elements from a global fit
to the data gives $\ov{\eta}=0.343\pm 0.015$ and leads to the SM
prediction \cite{ckmf}
\begin{eqnarray}
2\beta_s= 2.1^\circ \pm 0.1^\circ . \label{bsfit}
\end{eqnarray}
The combination of CDF, D0, ATLAS, and LHCb data on $B_s \to J/\psi
\phi$ and of LHCb data on  $B_s \to J/\psi \pi^+\pi^-$ gives \cite{hfag} 
\begin{eqnarray}
2\beta_s^{\rm exp}= 0.7^\circ \epm{5.2^\circ}{4.8^\circ} . \label{bsexp}
\end{eqnarray}
Of course, this average is dominated by the LHCb data. At this conference
Jeroen van Leerdam has reported $2\beta_s= 0.1^\circ \pm 4.8^\circ_{\rm
  stat} \pm 1.5^\circ_{\rm syst} $ from the combined LHCb analysis of
$B_s \to J/\psi K^+ K^-$ and $B_s \to J/\psi \pi^+\pi^-$ \cite{leerdam}.
In \eq{acpt2} an additional small contribution, the ``penguin pollution''  
has been neglected. This --presently uncalculable-- contribution
inflicts an additional error of order $1^\circ$ on $2\beta_s^{\rm exp}$
in \eq{bsexp}.

\subsection{Decay matrix $\Gamma_{12}^q$}
The calculation of $\Gamma_{12}^q$, $q=d,s$, is needed for the width
difference $\dg_q \simeq 2 |\Gamma_{12}^q| \cos \phi_q$ and the
semileptonic CP asymmetry $a_{\rm fs}^q =
\frac{|\Gamma_{12}^q|}{|M_{12}^{q}|} \sin \phi_q $. The theoretical 
results of Refs.~\cite{Beneke:1998sy,Beneke:2003az,Ciuchini:2003ww} have
recently been updated to \cite{ckm10} 
\begin{eqnarray}
\phi_s=0.22^\circ \pm 0.06^\circ, &&\qquad \mbox{~~~~~}\qquad 
\phi_d=-4.3^\circ \pm 1.4^\circ, \nn 
 a_{\rm fs}^{s} = (1.9 \pm 0.3) \cdot 10^{-5}, &&\qquad \mbox{and}\qquad 
 a_{\rm fs}^{d} = - (4.1 \pm 0.6) \cdot 10^{-4} . \label{phia}
\end{eqnarray}
The prediction of $\Gamma_{12}^q$ involves new operators in addition to
$Q$ in \eq{q}. However, the matrix element of $Q$ comes with the largest
coefficient, so that the ratio $\dg_q/\dm_q=|\Gamma_{12}^q|/|M_{12}^q|$
suffers from smaller hadronic uncertainties than $\dg_q$. Predicting
$\dg_q$ with the help of the experimental values $\dm_d^{\rm exp}= 0.507
\, \mbox{ps}^{-1}$ and $\dm_s^{\rm exp}$ in \eq{dmsexp} one infers from
Ref.~\cite{ckm10}:
\begin{eqnarray}
 \dg_d &=& \frac{\dg_d}{\dm_d} \dm_d^{\rm exp} = (27 \pm 5)\cdot 10^{-4}
 \, \mbox{ps}^{-1} \nn
 \dg_s &=& \frac{\dg_s}{\dm_s} \dm_s^{\rm exp} =  (0.090 \pm 0.018) 
 \, \mbox{ps}^{-1} \label{dgds}
\end{eqnarray}
The more recent update in Ref.~\cite{ckm12} differs from
Ref.~\cite{ckm10} in two respects: The quoted results are obtained in a
different renormalisation scheme and use the preliminary lattice results
of \cite{Bouchard:2011xj} instead of the averages from
\cite{ckmf}. Ref.~\cite{ckm12} finds for $\dg_s$:
\begin{eqnarray}
 \dg_s = \frac{\dg_s}{\dm_s} \dm_s^{\rm exp} =  (0.078 \pm 0.018) 
 \, \mbox{ps}^{-1}, \label{dgs}
\end{eqnarray} 
which is consistent with \eq{dgds}. In \eqsand{dgds}{dgs} all
uncertainties are added in quadrature, which may not be a 
conservative estimate of the overall error. The ranges comply with the 
LHCb measurement \cite{leerdam}   
\begin{eqnarray}
\dg_s^{\rm LHCb} = \lt[0.116\pm 0.018{}_{\rm stat}\pm 0.006_{\rm syst}\rt]
  \mbox{ps}^{-1}
\end{eqnarray}
and the world  average \cite{hfag}:
\begin{eqnarray}
\dg_s^{\rm exp} = \lt[ 0.089 \pm 0.012 \rt]  \mbox{ps}^{-1}.
\end{eqnarray}

\section{New physics in \boldmath \bbm}
\subsection{A model-independent analysis}
\begin{figure} 
\centerline{
\includegraphics[width=0.9\textwidth]{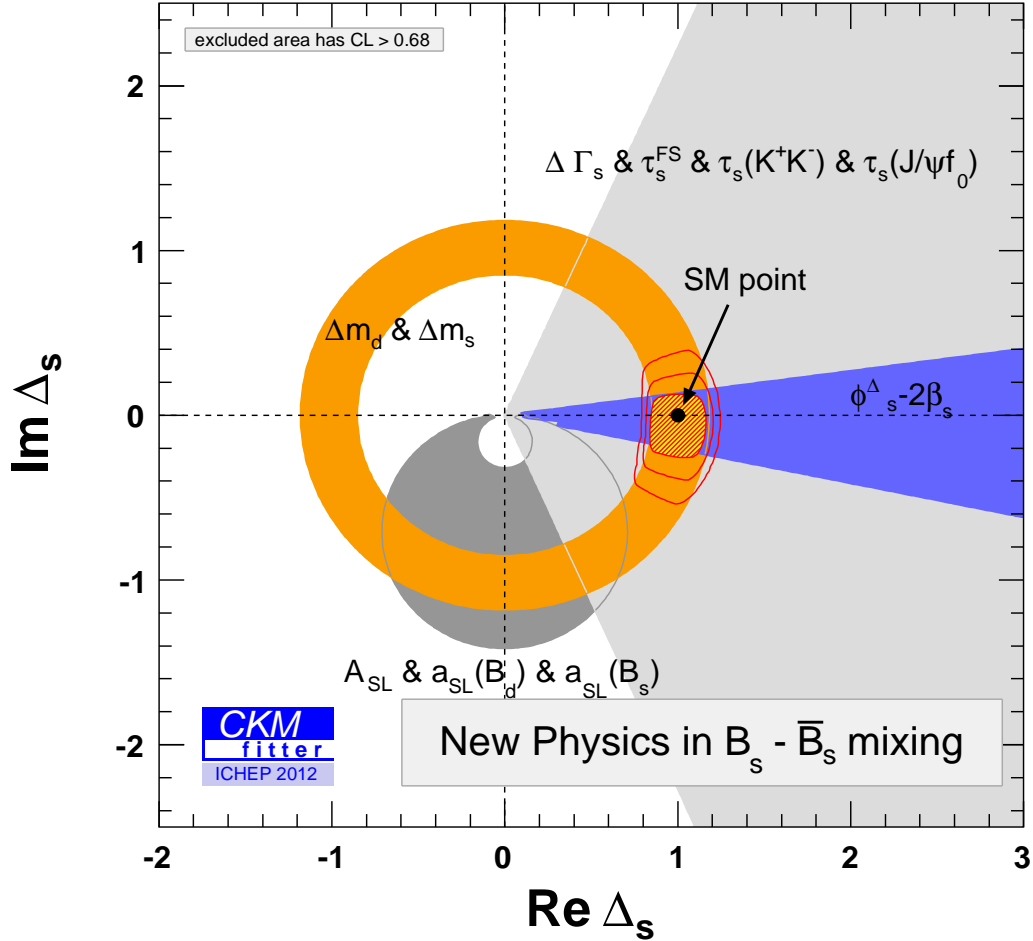}}
\caption{Allowed range for $\Delta_s$, see \cite{Lenz:2012az,ckmf} for details.
\label{fig:s}}
\end{figure}

\begin{figure} 
\centerline{
\includegraphics[width=0.9\textwidth]{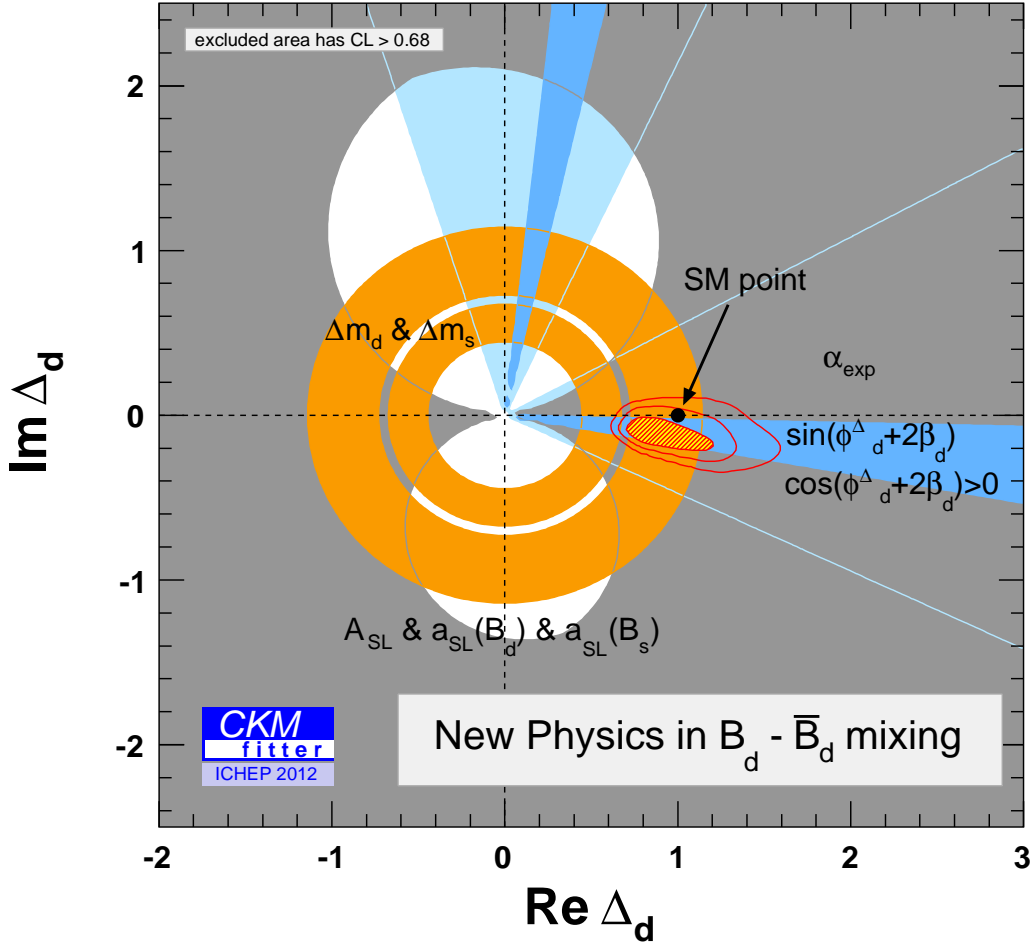}}
\caption{Allowed range for $\Delta_d$, see \cite{Lenz:2012az,ckmf} for details.
\label{fig:d}}
\end{figure}

If some NP amplitude adds to the \bbmq\ box diagram, both $|M_{12}^q|$
and $\phi_q$ may deviate from their SM predictions.  The D\O\ experiment
has measured \cite{Abazov:2010hv,Abazov:2010hj,Abazov:2011yk}
\begin{eqnarray}
A_{\rm SL}^{\rm D0} &=& 
 (0.532 \pm 0.039) a_{\rm fs}^d + (0.468 \pm 0.039) a_{\rm fs}^s\nn
&=& (-7.87 \pm 1.72  \pm 0.93 ) \cdot 10^{-3} , \label{dzero}
\end{eqnarray}
which is $3.9\sigma$ off the SM prediction inferred from \eq{phia},
\begin{eqnarray}
  A_{\rm SL} &=& (-0.20 \pm 0.03) \cdot 10^{-3}. 
\end{eqnarray}
Adding a NP contribution $\phi^{\Delta}_{d,s}$ to either $\phi_d$ or
$\phi_s$ in \eq{phia} may enhance $|A_{\rm SL}^{\rm D0}|$, but the very
same contribution will also affect the value of $\beta_s$ in \eq{bsexp}
and the value of $\beta = \arg =\arg(-V_{tb}V_{td}^{*}/ (V_{cb}
V_{cd}^{*} ))$ found from $a_{CP}(t)(B_d\to J/\psi K_S)$. One can
parametrise NP in \bbmd\ and \bbms\ by two complex parameters $\Delta_d$
and $\Delta_s$: 
\begin{eqnarray}
M_{12}^q & \equiv & M_{12}^{\rm SM,q} \cdot  \Delta_q \, ,
\qquad  \Delta_q \; \equiv \;  |\Delta_q| e^{i \phi^\Delta_q} .
\no
\end{eqnarray}
In summer 2010, before the advent of precision data from LHCb, a global
fit of all relevant flavour data to the CKM elements and $\Delta_d$
and $\Delta_s$ has resulted in a $3.6\sigma$ evidence of NP, with a
large negative NP phase $\phi^\Delta_s$ \cite{Lenz:2010gu}. In spring
2012 this fit has been repeated \cite{Lenz:2012az}; Figs.~\ref{fig:s} 
and \ref{fig:d} show an update of the results in Ref.~\cite{Lenz:2012az}
with the data of 2012 summer conferences \cite{ckmf}. 
The fit tries to accomodate $A_{\rm SL}$ in \eq{dzero} and a slightly
high value of the world average for $B(B\to \tau \nu)$ through
$\phi_d^\Delta<0$. Since $a_{CP}(t)(B_d\to J/\psi K_S)$ precisely fixes
$2\beta+\phi_d^\Delta=42.8^\circ \pm 1.6^\circ$, $\phi_d^\Delta<0$ comes
with a higher value of $\beta$ than in the SM.  ($\beta>\beta^{\rm SM}$
entails a larger value of $|V_{ub}|$ which governs $B(B\to \tau \nu)$.)
Contrary to the situation in 2010, the Standard Model point
${\Delta_s=\Delta_d=1}$ is merely disfavoured by 1 standard deviation,
consistent with natural statistical fluctuations. At the best-fit point
the problem with $A_{\rm SL}$ is only marginally alleviated.

One could relax the problem with $A_{\rm SL}$ without affecting
$a_{CP}(t)(B_d\to J/\psi K_S)$ and $a_{CP}(t)(B_s\to J/\psi \phi)$ by
postulating new physics in $\Gamma_{12}^d$ or $\Gamma_{12}^s$
\cite{Dighe:2010nj,Bobeth:2011st,Lenz:2012az}. Since $\Gamma_{12}^s$
originates from Cabibbo-favoured tree-level decays, it can hardly be
changed in a significant way without spoiling the value of the average
decay width $\Gamma_s=(\Gamma_H^s+\Gamma_L^s)/2$
\cite{Lenz:2010gu,Lenz:2012az}. The LHCb measurement $\Gamma_s^{\rm
  LHCb}=( 0.6580 \pm 0.0054 \pm 0.0066 )\,\mbox{ps}^{-1}$ \cite{leerdam}
implies
\begin{eqnarray}
\frac{\Gamma_d}{\Gamma_s} =
    \frac{\tau_{B_s}}{\tau_{B_d}} = 0.997 \pm 0.013 
\end{eqnarray}
in excellent agreement with the SM prediction $\tau_{B_s}/\tau_{B_d} =
0.998 \pm 0.003$ \cite{Keum:1998fd,ckm10}. NP in the doubly
Cabibbo-suppressed quantity $\Gamma_{12}^d$ is phenomenologically only
  poorly constrained, but requires a somewhat contrived model of NP.

\subsection{A supersymmetric SO(10) model}
The Minimal Supersymmetric Standard Model (MSSM) has many new sources of
flavour violation, which all reside in the supersymmetry-breaking
sector. It is no problem to get a big effect in a chosen FCNC process,
but rather to suppress big effects elsewhere. This \emph{supersymmetric
  flavour problem}\ is substantially alleviated with the lower bounds on
the squark massed placed by ATLAS and CMS. Grand Unified Theories (GUT)
offer the possibility to have ``controlled'' deviations from the CKM
pattern of flavour violation in the quark sector: In GUTs quarks and
leptons are unified in common symmetry multiplets, opening the
possibility to observe the large lepton-flavour mixing encoded in the
Pontecorvo-Maki-Nakagawa-Sakata (PMNS) matrix $U_{\rm PMNS}$ in quark
flavour physics \cite{Moroi:2000mr,Moroi:2000tk}. Consider { SU(5)}
multiplets:
\begin{displaymath}
        {
 {\bf \ov{5}_1} =
   \lt(\begin{array}{c} { d_R^c}\\ { d_R^c}\\ {d_R^c}\\
                e_L \\ -\nu_e \end{array} \rt), \qquad
 {\bf \ov{5}_2} =
   \lt(\begin{array}{c} { s_R^c}\\ { s_R^c}\\ {s_R^c}\\
                \mu_L \\ -\nu_\mu \end{array} \rt), \qquad
 {\bf \ov{5}_3} =
   \lt(\begin{array}{c} { b_R^c}\\ { b_R^c}\\ {b_R^c}\\
                \tau_L \\ -\nu_\tau \end{array} \rt). }
\end{displaymath}
If the observed large atmospheric neutrino mixing angle stems from a
rotation of $\bf \ov{5}_2$ and ${\bf \ov{5}_3}$ in flavour space, it
will induce a large $ \tilde b_R -\tilde s_R$-mixing. Contrary to the
situation with right-handed quark fields, rotations of right-handed
squark fields in flavour space are physical because of the soft
SUSY-breaking terms.  The Chang--Masiero--Murayama (CMM) model has
implemented this idea in a GUT based on the symmetry breaking chain
SO(10)$\to$ SU(5) $\to$ SU(3)$\times$SU(2)$_L\times$U(1)$_Y$
\cite{Chang:2002mq,Harnik:2002vs,Girrbach:2011an}.  In some weak basis
the Yukawa matrix of down (s)quarks is diagonalised as
\begin{displaymath}
  { \mathsf{Y}_d} = V_{\rm CKM}^*
\lt( \begin{array}{ccc}
y_d & 0& 0 \\
0 & y_s & 0 \\
0 & 0 & y_b
\end{array}\rt)
U_{\rm PMNS}
\end{displaymath}
and the right-handed down squark mass matrix has the following diagonal
form: 
\begin{align}
  \mathsf{m}^2_{\tilde d}\left(M_Z\right) & 
     = \mbox{diag}\, \left( m^2_{\tilde
      d}, \, m^2_{\tilde d}, \, m^2_{\tilde d} - \Delta_{\tilde d}
  \right) .\no
\end{align}
The real parameter $ \Delta_{\tilde d}$ is calculated from
renormalisation-group effects driven by the top-Yukawa coupling.

Rotating $ { \mathsf{Y}_d}$ to diagonal form puts the large
atmospheric neutrino mixing angle into $  \mathsf{m}^2_{\tilde d}$:
\begin{align}
   U_{\rm PMNS}^\dagger \, \mathsf{m}^2_{\tilde d}\, U_{\rm PMNS}&  =
  \begin{pmatrix}
    m^2_{\tilde d} & 0 & 0 \\ 
    0 & m^2_{\tilde d} - \frac{1}{2}\,
    \Delta_{\tilde d} & - \frac{1}{2}\, \Delta_{\tilde d}\,
   e^{i\xi} \\ 0 & - \frac{1}{2}\, \Delta_{\tilde d}\,    
   e^{-i\xi} & m^2_{\tilde d} - \frac{1}{2}\, \Delta_{\tilde d}
  \end{pmatrix} \label{trib}
\end{align}
The CP phase $\xi$ affects CP violation in \bbms !

In Ref.~\cite{Girrbach:2011an} we have confronted the CMM model with
flavour data. The analysis involves seven parameters, which we have
fixed by choosing values for the squark masses $ M_{\tilde u}$, $
M_{\tilde d}$ of right-handed { up} and { down squarks}, the
trilinear term $ a_1^d$ of the first generation, the gluino mass $
m_{\tilde g_3}$, $\tan\beta$, and the sign of the higgsino mass parameter
$\mu$. We have considered { \bbms}, $ b\to s\gamma$, $
\tau \to \mu \gamma$, { vacuum stability bounds}, lower bounds on
sparticle masses and the lower bound on the lightest Higgs boson. 
From these inputs first universal SUSY-breaking terms defined at a fundamental
scale near the Planck scale have been determined through the
renormalisation group equations (RGE). Subsequently we have used the RGE  
to determine all low-energy parameters, with the MSSM as the low-energy
theory. 

Two experimental results of the year 2012 put the CMM model under
pressure: First, the sizable neutrino mixing angle $\theta_{13}$ leads
to an unduly large effect in $B(\mu\to e\gamma)$. In \eq{trib} I have
tacitly assumed a $U_{\rm PMNS}$ with tri-bimaximal mixing,
corresponding to $\theta_{13}=0$. For the actual value
$\theta_{13}\approx 8^\circ $ the $(1,2)$ element of the charged slepton
mass matrix gets large, and one has to resort to much larger sfermion
masses than those considered in Ref.~\cite{Girrbach:2011an}. Second, the
Higgs mass of 126$\,\gev$ challenges the model and we are unable to find
parameters which simultaneously satisfy the Higgs mass constraint and
the experimental upper bound on $B(\mu\to e\gamma)$ \cite{ns}.  The
problem with the Higgs mass could be circumvented by considering the
NMSSM as the low-energy theory.

\section{Conclusions} 
LHCb has provided us with a significantly better insight into the \bbms\
complex. $\dm_s$ and $\dg_s$ comply with the SM, but we need better
lattice data for the hadronic matrix elements involved. Theoretical
uncertainties still permit an ${\cal O} (20\%)$ NP contribution to the
\bbs\ amplitude $M_{12}^s$.  While in 2010 the D\O\ result for $A_{\rm
  SL}$ could be explained in scenarios with NP only in $M_{12}^{d,s}$,
the LHCb data on $B_s \to J/\psi \phi$ now prohibit this solution. An
alternative explanation invoking new physics in $\Gamma_{12}^s$ is not
viable, because this will spoil the ratio $\Gamma_s/\Gamma_d$ which
agrees well with the SM prediction. Maybe it is worthwile to look at NP
in $\Gamma_{12}^d$, although this possibility leads to somewhat
contrived models. Models of GUT flavour physics with $\widetilde b \to
\widetilde s$ transition driven by the atmospheric neutrino mixing angle
are under pressure from the large value of $\theta_{13}$, which induces
a too large $B(\mu\to e\gamma)$. Moreover, in the studied CMM model it
seems impossible to accomodate the measured value of the lightest Higgs
mass, if one insists on the MSSM as the low-energy theory. 
  
\section*{Acknowledgments} 
I am grateful for the invitation to this stimulating and enjoyable
conference! The work presented in this talk is supported by 
BMBF grant 05H12VKF. 

\section{References}

\bibliography{discrete12}
\bibliographystyle{iopart-num}

\end{document}